\documentclass[global,twocolumn]{svjour}

\usepackage{amsmath,amssymb}
\usepackage{graphics}

\journalname{Appl. Phys. B}

\begin{document}

\title{Controlling mode locking in optical ring cavities}
\author{G. Krenz \and S. Bux \and S. Slama \and C. Zimmermann \and Ph.W. Courteille}

\institute{Physikalisches Institut, Eberhard-Karls-Universit\"at T\"ubingen,
\\Auf der Morgenstelle 14, D-72076 T\"ubingen, Germany}

\date{Received: date / Revised version: date}

\maketitle

\begin{abstract}
Imperfections in the surface of intracavity elements of an optical ring resonator can scatter light from one mode into the counterpropagating mode. The phase-locking of the cavity modes induced by this backscattering is a well-known example that notoriously afflicts laser gyroscopes and similar active systems. We experimentally show how backscattering can be circumvented in a unidirectionally operated ring cavity either by an appropriate choice of the resonant cavity mode or by active feedback control.


\end{abstract}

\section{Introduction}
\label{intro}

Optical ring cavities have very interesting particularities distinguishing them from ordinary linear cavities. The most important one is that the counterpropagating modes have independent photon budgets, i.e.~a light scattering object can pump photons from one mode into the reverse mode. The phase of the standing wave formed by interference of the counterpropagating modes then becomes a degree of freedom. This features is exploited in ring laser gyroscopes, whose central device is a ring cavity. When the ring cavity rotates about the normal vector of the plane in which the laser beams propagate, the resonant frequencies of the counterpropagating optical modes of the ring cavity are Doppler-shifted by an amount proportional to the rotation velocity. This phenomenon is known as the Sagnac effect. A well-known problem of gyroscopes is mode locking, i.e.~an undesired complete phase synchronization of the counterpropagating waves at low rotation velocities \cite{Aranowitz66,Chow85,Scully86,Faucheux88}, when the resonant frequencies of the waves are almost degenerate. 

The problem arises when some amount of light in a ring cavity mode is scattered into the solid angle of the counterpropagating mode. This light seeds the cavity mode into which it is scattered and thus forces it to adopt the same frequency. Scattering from mirror imperfections is almost unavoidable. In practice in a laser gyroscope, this phase locking is circumvented by lifting the mode degeneracy by various means, i.e.~Faraday effect, using higher-order transversal modes or modulation of the mirrors' positions. In particular, dithering techniques where the position of one or several mirrors is slightly modulated and the signal phase-sensitively analyzed have led to amazing resolutions in state-of-the-art laser gyros. E.g.~the tumbling of the earth's rotation axis can be measured with a precision of 1 part in $10^8$ \cite{Stedman01,Schreiber04}

Mode locking is always a problem when the phase of the standing wave in the ring cavity is needed as a degree of freedom. This is not only the case in laser gyros, but also when the subject of interest is the interaction of a medium inserted into the cavity with the counterpropagating optical modes. Recently, ring cavities have been rediscovered for cold atom optics \cite{Kruse03,Nagorny03}. Atoms suspended within the mode volume of the cavity scatter light from the pumped into the reverse mode thus establishing a phase relation between the modes. However the atoms, although they have some inertia, cannot be considered immobile. From this results a self-consistent dynamics with a time-dependent frequency shift between the locked cavity modes \cite{Kruse03b}. The phenomenon has been termed the collective atomic recoil laser (CARL). In such a system additional backscattering from the mirror surfaces plays a noticeable role leading, under certain circumstances, to degenerate frequency locking for a atomic coupling strength below a certain threshold \cite{Slama06}.

The problem is particularly disturbing when the coupling mechanism under investigation is weak. This is the case in the superradiant regime of the CARL \cite{Slama06}, which arises when the collective coupling is weak compared to the cavity decay rate. It is even more prominent in the so-called quantum regime \cite{Piovella01b}. In the present paper, we experimentally study mirror backscattering in a high-finesse ring cavity. Most of the solutions developed to reduce this problem in laser gyros with an active medium are not applicable to our system. Here we propose a scheme which can be adapted to empty Sagnac interferometer, such as our ring cavity, pumped from only one side. It consists in compensating mirror backscattering by injection of an additional light field, whose phase is controlled via an active feedback mechanism. In contrast to the gyro, we want the absence of any light field in one mode to avoid seeding.

\section{The experiment}
\label{sec:1}

The ring cavity is very similar to the one used in Refs.~\cite{Kruse03,Kruse03b}. It consists of one plane incoupling mirror with the intensity transmission $T_1=11\times10^{-6}$ and two curved high reflecting mirrors with $T_2=T_3=1.5\times10^{-6}$. The cavity is $L=8.7~$cm long and has a beam waist of $w_0=107~\mu$m. For the sake of definiteness, we describe in the following the cavity modes by their field amplitudes $\alpha_{\pm}$ scaled to the field per photon, so that $|\alpha_{\pm}|^2$ is the number of photons in the modes \cite{Gangl00}. Only the mode $\alpha_+$ is continuously pumped by a titanium-sapphire laser at $\lambda=797~$nm. The laser can be stabilized to this mode using the Pound-Drever-Hall method. The counterpropagating (probe) mode is labeled by $\alpha_-$. The ring cavity can be operated in two ways depending on the polarization of the incoupled light. For $p$-polarized light a finesse of 87000 is determined from a measured intensity decay time of $\tau\approx3.8~\mu$s. For $s$-polarized light the finesse is 6400.
	\begin{figure}[ht]
		\centerline{\scalebox{0.67}{\includegraphics{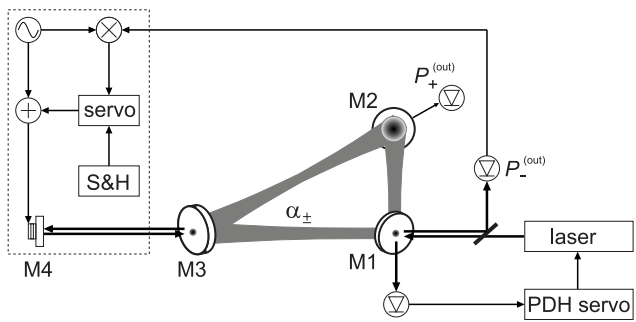}}}\caption{
			Schematic view of the ring cavity (mirrors M1, M2 and M3). It is pumped from one side by a titanium-sapphire laser, which is
			phase-locked to an eigenfrequency of the cavity. The light in the pumped mode ($\alpha_+$) and the reverse mode ($\alpha_-$)
			leaking through a high reflecting mirror of the cavity is monitored. 
			For some experiments, the setup is extended by the components shown within the dashed box. A fraction of the light field
			$\alpha_+$ leaking through mirror M3 is fed back into the mode $\alpha_-$. Its phase and amplitude are controlled by a servo
			loop such as to cancel the field due to backscattering of the mode $\alpha_+$. The servo can be interrupted by a
			sample-and-hold (S\&H) circuitry.}
		\label{Fig1}
	\end{figure}

\begin{figure}[ht]
		\centerline{\scalebox{0.47}{\includegraphics{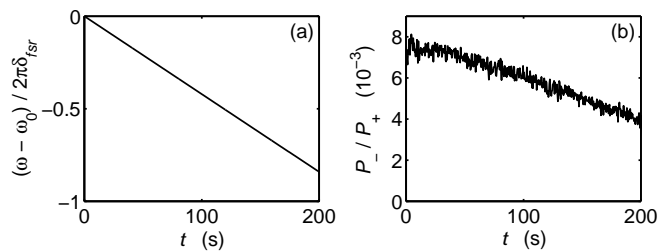}}}\caption{
			(a) Example of an observed time-evolution of the resonant frequency of the cavity, when the cavity is exposed to temperature
			variations. The frequency $\omega$ of the pump laser locked to the cavity drifts away from its initial value $\omega_0$. 
			(b) Simultaneously monitored fraction of power in the reverse mode $P_-/P_+$. The power in the pumped mode is $P_+=1$~W.}
		\label{Fig2}
	\end{figure}

We measure the intracavity light power via the fields leaking through the cavity mirrors (see Fig.~\ref{Fig1}). The outcoupled light power is related to the intracavity power by $P_-^{(out)}=T_1P_- =T_1\hbar\omega\delta_{fsr}~|\alpha_-|^2$ and $P_+^{(out)}=T_2P_+$ respectively, where $\delta_{fsr}=(3.4\pm0.05)~$GHz is the free spectral range of the cavity and $\omega/2\pi$ the laser frequency. Although no light is injected into the reverse mode $\alpha_-$, we observe that it carries a considerable amount of light, between $0$ and $1\%$ of the light power in the pumped mode. Furthermore, thermal drifts which slightly influence the cavity geometry [and hence the resonant frequency as shown in Fig.~\ref{Fig2}(a)] also cause variations of the reverse power [see Fig.~\ref{Fig2}(b)].

The origin of the light found in the reverse mode is backscattering from the surfaces of the three cavity mirrors. The mirrors (Research Electro-Optics, Inc.) generate losses due to absorption in the dielectric layers or to scattering from imperfections at the surfaces. The losses due to scattering are typically below $S=10$~ppm for high-quality surface mirrors \cite{Faucheux88}. Let us assume that the light scattering from a mirror surface is isotropic. The amount of light scattered into the reverse mode is proportional to the solid angle \cite{Kruse04}
\begin{equation}\label{Eq01}
	\frac{\Omega_s}{4\pi} = \left(\frac{w_0}{4L}\right)^2~.
\end{equation}
With the amplitude reflection coefficient $\beta=\sqrt{S\Omega_s/4\pi}$ the intermode coupling strength reads
\begin{equation}\label{Eq02}
	U = \pi\delta_{fsr}\beta~.
\end{equation}
Therefore, in units of the cavity decay rate $\kappa=(2\tau)^{-1}$ the coupling strength is $U\simeq0.06\kappa$.

\section{Frequency-dependence of mirror backscattering}
\label{sec:2}

Figs.~\ref{Fig2}(a,b) suggest a correlation between the drifts of the cavity length and the amount of backscattering. Information about the nature of the scatterers can only be gathered by injecting light into the cavity, which is only possible when the pump laser is resonant with a cavity mode. Apart from tuning the resonant frequency of the cavity by modifying its length, one can probe the discrete set of longitudinal modes by varying the pump laser frequency. The analysis of the reverse power as a function of the pump frequency should lead to a better understanding of the temporal behavior of the reverse power. 
	\begin{figure}[ht]
		\centerline{\scalebox{0.49}{\includegraphics{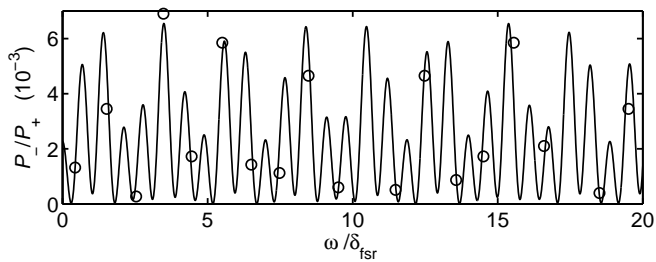}}}\caption{
			The circles denote the ratio between backscattered and pump light measured as a function of the longitudinal mode number in the 
			vicinity of $N=110695$. The solid line is a fit of the frequency dependence of the mirror backscattering calculated from
			expression~(\ref{Eq02}). The curve is physically meaningful only at the locations where the periodic boundary is fulfilled, i.e.~at
			integer values of $\omega/\delta_{fsr}$. The location of the scatterers are assumed to coincide with the mirror positions, $r_1=0$,
			$r_2=-2.58~$cm, and $r_3=3.65~$cm. The fit yields the mode coupling strengths of the three mirrors $U_1=0.034\kappa$, $U_2=0.017\kappa$,
			and $U_3=0.031\kappa$, and the frequency offset $\omega_0$.}
		\label{Fig3}
	\end{figure}

The experiment exhibited in Fig.~\ref{Fig3} indeed shows a strong dependence of the backscattering rate on the frequency of the pump laser, when it is resonant to an eigenfrequency of the cavity. A very simple model can be used to describe the transfer of light between the modes by the backscattering mechanism. For simplicity we assume that only one mode, $\alpha_+$, is pumped by resonant light. The pump laser is tightly locked to this mode. The losses due to transmission through the cavity mirrors or to backscattering at the mirror surfaces are negligible, so that the mode $\alpha_+$ is in a steady state.

At a given frequency, the mode coupling induced by mirror backscattering can be expressed by a single complex quantity: The coupling strength $B$ describes at which rate a scatterer sitting at a given location shuffles photons from one mode into the reverse mode. $k=\omega/c$ is the wavenumber of the resonant pump light. The basic equation for the evolution of the reverse mode $\alpha_-=\alpha_-(t)$ can then be written as \cite{Kruse04},
\begin{equation}\label{Eq03}
	\dot{\alpha}_- = -\kappa\alpha_--iB\alpha_+~.
\end{equation}
The model is lend from a more general system \cite{Gangl00,Kruse03b}, where the backscattering is provided by atoms. In contrast, here we assume spatially fixed scatterers. The cavity decay rate $\kappa$ describes the losses for the reverse mode. Obviously, the coupling strength is frequency-dependent $B=B(\omega)$. Since the cavity consists of three mirrors, the scatterers are spatially separated. If we assume the presence of one microscopic scatterer sitting on each mirror surface localized at the position $r_n$ along the optical axis, we get,
\begin{equation}\label{Eq04}
	B = \sum\nolimits_{n=1}^3U_ne^{2ikr_n}~.
\end{equation}
The reverse power results from interference of the waves backscattered from all three cavity mirrors. The stationary solution of the CARL equations~(\ref{Eq01}) without atoms, but with three immobile point-like scatterers is
\begin{equation}\label{Eq05}
	\frac{P_-}{P_+} = \frac{|\alpha_-|^2}{|\alpha_+|^2} = \frac{|B|^2}{\kappa^2}~.
\end{equation}
This expression is invariant upon simultaneous translations of all mirrors, $r_n\rightarrow r_n+a$. Periodic boundary conditions are satisfied when the cavity is resonant, i.e.~$r_n\rightarrow r_n+L$ holds separately for every $n$, but only if $\omega=N\delta_{fsr}$. Fitting the expression to the measured values (see Fig.~\ref{Fig3}) permits to determine the contributions $U_n$ of every mirror to backscattering. The values are consistent with the rough estimation (\ref{Eq02}) from the mirror surface roughness.

\bigskip

Based on this model we may also understand the observation of Fig.~\ref{Fig2}(b). A strictly uniform expansion of the cavity length by $\Delta L$ modifies the distances between any two mirrors by $(r_m-r_n)\Delta L/L$, but leaves the quantity $B$ unchanged. The observations of Fig.~\ref{Fig2}(b) thus result from a \textit{non-uniform} expansion of the cavity. 

The assumption that the scattering defects or dust particles at the mirror surfaces are point-like is a course approximation. In reality scatterers sitting on the same mirror may be much larger than an optical wavelength and cross many phase planes. If the scattering is not uniform across the scatterer, the dispersion spectrum $B(\omega)$ is additionally modulated. However, since the spatial distribution of the scatterer being on the order $w\cos{\alpha}$, where $\alpha=22.5^{\circ}$ and $45^{\circ}$ are the reflection angles in the ring cavity, the spectral width will be large, $\Delta\omega/2\pi\approx c/w\simeq100\delta_{fsr}$. Therefore, the broadening of the scatterers distribution on one mirror only slightly influences the spectrum recorded in Fig.~\ref{Fig3}. Nevertheless, it may account (together with thermal drifts during the data recording) for deviations between the measurement and the fitted curve. From the experimental point of view, the most interesting feature is that backscattering can be reduced by a proper choice of the resonant cavity mode.

\section{Annihilation of mirror backscattering}
\label{sec:3}

We have seen in the previous section that the frequency-dependence of the amount of backscattered light is due to interference of the light reflected from all three cavity mirrors. If a forth scatterer could be introduced on purpose, its amplitude and phase could be designed such as to cancel out the field produced by the cavity mirrors. Such an additional scatterer can be simulated by injecting a laser beam into the reverse mode of the ring cavity. Moreover, by continuously monitoring the amount of light in the reverse mode and using this information to control the phase of the injected beam, the intensity in the reverse mode can be stabilized to zero by a servo loop as shown in Fig.~\ref{Fig1}. This is necessary since, as shown in Fig.~\ref{Fig2}, the amount of backscattering varies in time due to thermal drifts. The bandwidth of the servo loop must be larger than the drift rate of the amount of backscattering, which is on the time scale of minutes. However, one has to be aware that the phase of the injected light field may be subject to acoustic noise itself. Nevertheless, a piezo transducer (mounted to mirror M4) appears to be sufficient for applying the corrections.

Concretely, we feed back part of the light leaking through mirror M3 (see Fig.~\ref{Fig1}), thus providing an additional tunable backscattering mechanism. Let us denote the coupling strength generated by the feedback interferometer by $U_0$. Then taking the mirror backscattering from Eq.~(\ref{Eq04}), including the injected light field $U_0e^{2ik\left[r_0+\Delta r_0\cos(\Omega t)\right]}$, the intracavity power ratio is
\begin{equation}\label{Eq05}
	\frac{P_-}{P_+} = \frac{\left|B-U_0e^{2ik\left[r_0+\Delta r_0\cos(\Omega t)\right]}\right|^2}{\kappa^2}~.
\end{equation}
Using a small modulation excursion $k\Delta r_0$, the demodulated signal, $D=\operatorname{Re}\left[P_-^{(out)}e^{i\Omega t+i\theta}\right]$, is
\begin{equation}\label{Eq06}
	D \propto \operatorname{Re}\left[e^{i\Omega t+i\theta}\left|B-U_0e^{2ikr_0}\left(1+2ik\Delta r_0\cos(\Omega t)\right)\right|^2\right]~.
\end{equation}
Setting $B\equiv|B|e^{2ikr_s}$ the only contribution surviving temporal integration (i.e.~low-pass filtering with time constant $\tau$), $\bar{D}\equiv\frac{1}{\tau}\int_0^{\tau}Ddt$, is
\begin{equation}\label{Eq07}
	\bar{D}\propto 2k\Delta r_0U_0|B|\cos\theta\sin\left[2k\left(r_0-r_s\right)\right]~.
\end{equation}
Choosing $\theta=0$ we obtain an error signal crossing 0 around $r_0=r_s$. When the phase of the injected beam is locked, to null the power in the reverse mode, $\kappa^{-2}P_+\left(U_0-|B|\right)^2$, the injected field amplitude must be tuned to satisfy $U_0=|B|$. 
	\begin{figure}[ht]
		\centerline{\scalebox{0.48}{\includegraphics{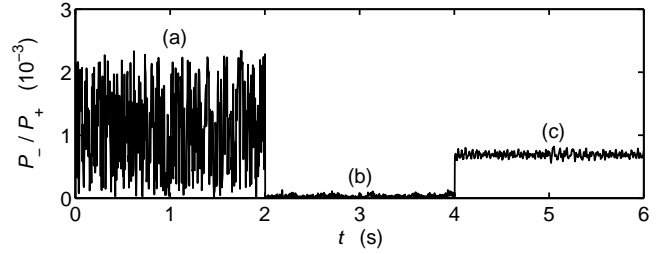}}}\caption{
			(a) Temporal evolution of the power in the reverse mode when a light beam is injected into the reverse via an additional mirror
			M4, as shown in Fig.~\ref{Fig4}.
			(b) Reduction of the intensity fluctuations by servo locking the phase of the injected light beam. The position of mirror M4 is
			modulated with the frequency $\Omega/2\pi=1~$kHz.
			(c) The injection light beam is blocked.}
		\label{Fig4}
	\end{figure}

\bigskip

As a proof of principle, we set up the locking scheme Fig.~\ref{Fig1}. We reinject a light beam leaking out of the pump mode through one of the cavity mirrors into the reverse mode. The power in the reverse mode, whose behavior is shown in part (a) of Fig.~\ref{Fig4}, exhibits strong fluctuations due to acoustic noise randomly shifting the phase of the injected beam. The power fluctuations vary within the range $\kappa^{-2}P_+\left(U_0\pm|B|\right)^2$. 

By controlling the phase of the reinjected beam by a servo loop its light can be made to cancel out the light in the reverse mode due to mirror backscattering. Part (b) of  Fig.~\ref{Fig4} shows the power in the reverse mode when the servo loop is operating. In part (c) the injection beam is blocked, so that the power in the reverse mode is due to mirror backscattering only. The backscattering ratio is $P_-/P_+=0.07\%$. Comparison of (b) and (c) reveals a more than 10-fold reverse power reduction. 

For detecting the coupling dynamics due to the medium inserted into the ring cavity, the active servo must be interrupted by a sample-and-hold circuitry. Otherwise, the servo loop works to compensate the atomic backscattering mechanism under investigation. Since the typical time scale for the medium-induced coupling is $\mu$s, the inertia of the mirrors' positions is sufficient to guarantee steady-state conditions during the measurement, even when the servo loop is briefly interrupted.

\section{Conclusion}
\label{conclu}

Mirror backscattering in a ring cavity is not a doom. For a fixed cavity geometry it can be minimized by tuning the laser to a longitudinal resonator mode where backscattering happens to be small. This is however not possible if the cavity is subject to non-uniform drifts. In this work we discussed and demonstrated how to couple an additional light field into the reverse mode with an amplitude and phase chosen such as to obtain destructive interference \cite{Note1}. This measure allows us to limit the power in the reverse mode to below 200~ppm of the pump power. A further improvement may be expected from a more stable mechanical mounting and by replacing the piezo modulation with an electro-optic phase modulation at high frequencies. The optical sidebands generated by such a phase modulation could be used to implement a Pound-Drever-Hall type locking scheme, whose advantage is a very large servo bandwidth.

To give a rough idea of how much the reverse power can be suppressed with a better servo loop, we estimate the shot noise limit. The number of photons recorded by the photodetector of the locking servo within the $\tau\simeq1~$ms integration time is $n_-^{(out)}=T_1\delta_{fsr}\tau n_-$. This signal is subject to Poisson-distributed noise, $\Delta n_-^{(out)}=\sqrt{n_-^{(out)}}$. The shot noise level is reached when the phase $\phi_-$ of the injected laser beam controlled by the servo has fluctuations of only $\Delta\phi_-=1/\Delta n_-^{(out)}$. These phase excursions produce a reverse power ratio of
\begin{equation}\label{Eq08}
	\frac{\Delta P_-}{P_+} = \frac{\left|B-U_0e^{2ikr_0+i\Delta\phi_-}\right|^2}{\kappa^2}
	\approx \frac{|U_0|^2\Delta\phi_-^2}{\kappa^2}~,
\end{equation}
when the locking servo regulates to minimum reverse power and $|B|\approx U_0$. Hence,
\begin{equation}\label{Eq09}
	\frac{\Delta P_-}{P_+} \approx \frac{|B|^2}{\kappa^2T_1\delta_{fsr}\tau n_-} = \frac{1}{T_1\delta_{fsr}\tau n_+}~.
\end{equation}
With typical pump powers of $P_+\simeq1~$W and under the premise of a perfectly operating servo system the shot noise level is $\Delta P_-/P_+\simeq3\times10^{-11}$. This leaves room for improvement using a faster servo loop, e.g.~having a much larger servo bandwidth. 

The minimization of unwanted backscattering will be crucial for the detection of weak coupling forces, such as they occur in the small gain limit of the collective atomic recoil laser \cite{Slama06}.

\bigskip

This work has been supported by the Deutsche Forschungsgemeinschaft (DFG) under Contract No. Co~229/3-1.

\end{document}